# Inventions on soft keyboards
## -A TRIZ based analysis

**Umakant Mishra**

Bangalore, India

umakant@trizsite.tk

http://umakant.trizsite.tk

## Contents

## 1. Introduction

The soft keyboards are onscreen representation of physical keyboard having alphanumeric characters and other controls. The user operates the soft keyboard with the mouse, a stylus or other pointing device. The soft keys don't have any mechanical component.

The soft keyboards are used in many public places for informational purpose, educational systems and financial transactional systems. A soft keyboard is convenient in some cases where a hard keyboard is difficult to manage. The soft keyboard is a substitute of a physical keyboard and is displayed on the screen. It displays the same type of alphanumeric and control keys like the keys on the actual keyboard.

There are many inventions on a soft keyboard which makes the soft keyboard more efficient and effective.

## 2. Inventions on soft keyboard

### 2.1 Hexagonal soft keyboard (Patent 5805157)

**Background problem**

The traditional shape of the soft keys is rectangular which is not very effective. The user normally uses the central portion of the rectangle and not the corners. Besides, the user does not like to cover the label of the key while clicking or touching the key and normally hit at the bottom of the key. This sometimes makes him to hit the key lying below the desired key. How to solve this problem?

**Solution provided by the invention**

Bertram et al. invented this hexagonal shape soft keyboard (patent 5805157, assigned to IBM, Issued in Sep 98) this hexagonal shape of soft keyboard. Hexagonal cells enable display of larger areas for engagement by the finger or other input pointer used by a user. A user can be permitted a choice between hexagonal and rectangular cells.

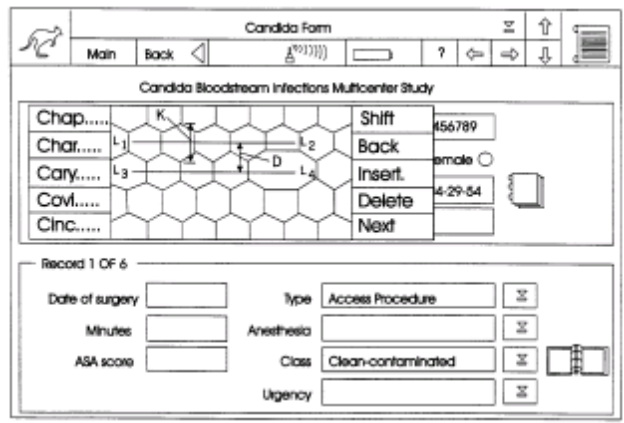

**TRIZ based analysis**

The invention makes the shape of the soft key approximately a circle or ellipse as closely as possible **(Principle-14: Curve)**.

It is desirable to interlock the shapes so that the spaces between keys are not wasted. In that case a hexagon is the right shape **(Principle-16: Partial or Excessive Action)**

### 2.2 Computer programmed soft keyboard system (5818451)

**Background problem**

The problem with the conventional shape of the keys of the soft keyboard is that the users tend to target the lower portion of the keys with their fingers, input stylus or mouse pointer instead of targeting the key's center. This leads to cases of user accidentally selecting the key below the one he actually intended. Again because of the thickness of the monitor glass the user may make more mistakes, as the user's line of sight may not be perpendicular to the two surfaces. It is therefore necessary to improve the accuracy of a soft keyboard by adjusting the interpretation of the sensed input to more accurately reflect what the user intends.

**Solution provided by the invention**

Bertram et al. disclosed this method of touch screen input area (patent 5818451, Assigned to IBM, Issued in Oct 1998) having a provision of an offset adjustment of the actually sensed touch input point to properly interpret the likely desired selection or area of contact with regard to what will be the most probably area of contact on the soft keyboard. The offset adjustment enables improved accuracy of data entry a user systems embodying this invention.

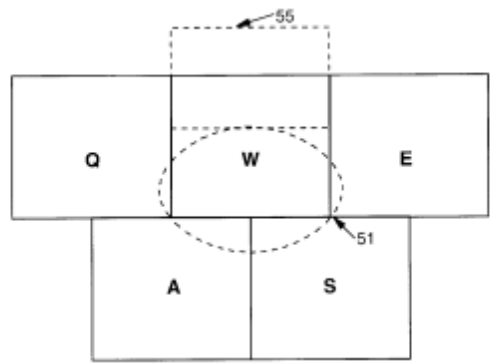

**TRIZ based solution**

The system should automatically detect and select the key that the user intends to hit **(Ideal Final Result)**.

The invention, looking at the trend of users' contact points, defines a more probable area of contact than the actual boundary of the soft key which may even overlap to some portion of the adjacent keys **(Principle-16: Partial and Excessive Action)**.

The method makes the contact points closer to circles and definitely not rectangular **(Principle-14: Curve)**.

**2.3 Enhancement of Soft keyboard (Patent 5963671)**

**Background problem**

A soft keyboard is implemented by displaying an image of the keyboard, which the user interacts through a pointing device like mouse or stylus. However data entry is quite slow in a soft keyboard as the user must first locate the desired key and then move the pointer to that key. This kind of keyboards is not quite suitable for applications that require a substantial amount of data entry. How to make a soft keyboard more efficient?

**Solution provided by the invention**

The problem is solved by US patent 5479536 (invented by Comerford, assigned to IBM in Issued in Dec 95) which provides the solutions as above. The soft keyboard uses a trigram data to display only those characters which are most likely to be entered by the user. The characters are highlighted for easy identification. This feature makes it convenient to choose the one desired out of the list.

Later US patent 5963671 (invented by Comerford, et al., assigned to IBM in Oct 99) further enhances the soft keyboard emphasizing the graphic presentation of the most likely full words or keys for quick selection. This invention also brings the mouse pointer closer to the most likely letter selection to speed up the entry.

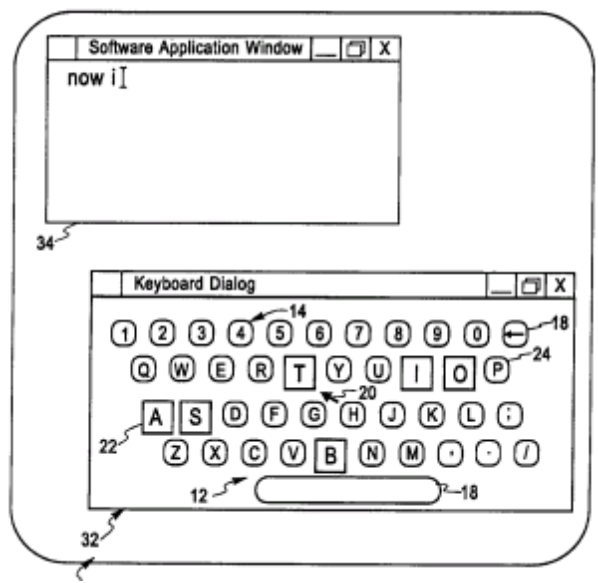

**TRIZ based analysis**

It's difficult to locate the keys on a soft keyboard **(Original problem)**.

The keyboard should locate the desired key for us **(Ideal Final Result)**.

According to the former invention the computer makes an intelligent guess and locate a few most likely keys by referring to a dictionary reference table to prompt the next character **(Principle-10: Prior Action)**.

The computer also displays those keys in bright or in a different color to be identified easily **(Principle-32: Color Change)**.

The later invention moves the pointer closer to the most likely key **(Principle-10: Prior Action)**.

### 2.4 User defined keyboard entry system (Patent 5936614)

**Background problem**

The touch sensitive keyboards are displayed on the main display and operated through touch sensors. In some cases the system provides a mechanism of adjusting display of the keyboard but that may not suite the positioning of application screens. It would be nice to enable the user to develop his own customized keyboard without writing codes and without loosing compatibility with existing applications.

**Solution provided by the invention**

Larry An et al. disclosed a mechanism (patent 5936614, Assigned to IBM, Aug 99) that enabled a user to design touch-activated keyboards for use on a display surface without writing code. The computer system enables the user to resize and move the keyboard image anywhere on the display to allow the concurrent display of other applications running in an integrated operating environment. Computer system actions can be assigned by the user to each key. An execution unit displays the selected keyboard, determines whether a key has been touched and executes the appropriate action for each key touch.

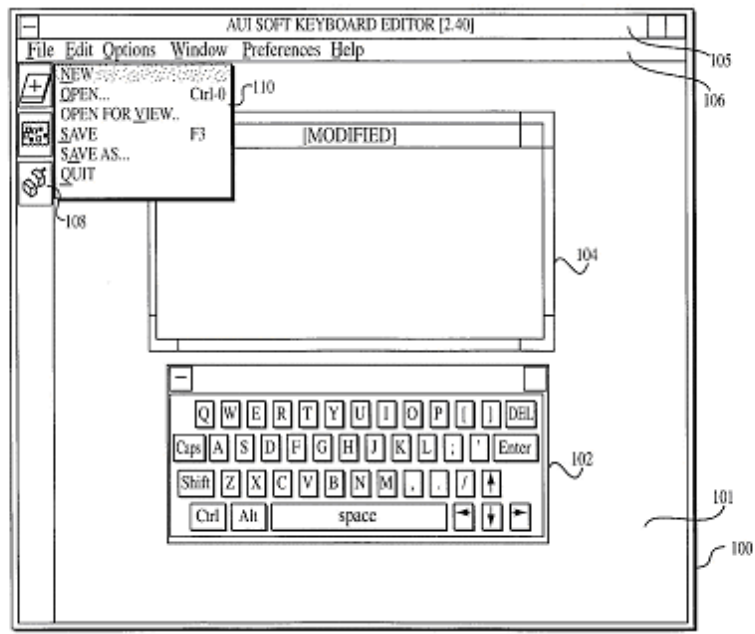

**TRIZ based analysis**

The invention provides reorganizing features to the touch sensitive keyboard **(Principle-15: Dynamize)**.

## 2.5 Predictive keyboard (Patent 6573844)

**Background problem**

A PDA has a very small keyboard and not convenient for typing. Often there is a preference to use the soft keyboards, which displays a keyboard on the screen and guides the users for typing. A disadvantage of the soft keyboard is again the size which leads to typing errors.

**Solution provided by the invention**

Venolia et el. disclosed a predictive keyboard (Patent Number 6573844, Assigned to Microsoft Corporation, June 2003).

The method displays a soft keyboard where the predicted keys are displayed on the soft keyboard differently than the other keys on the keyboard. For example, the predicted keys may be larger in size on the soft keyboard as compared to the other keys. This makes the predicted keys more easily typed by a user as compared to the other keys.

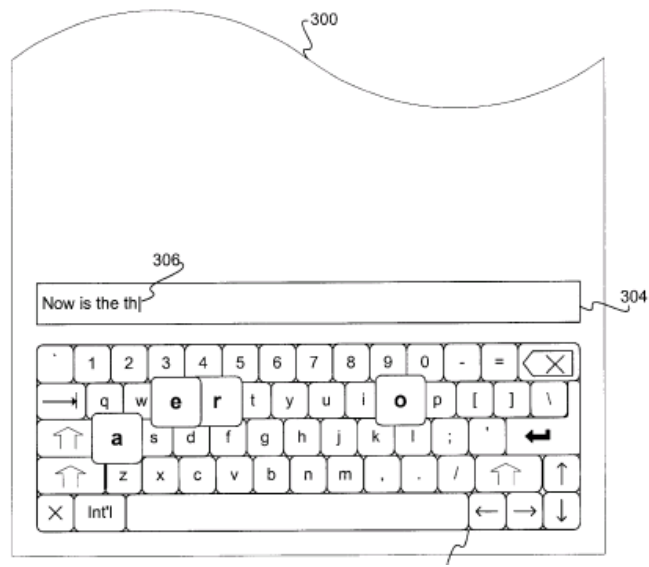

**TRIZ based analysis**

The keys should be small to fit in the display screen, but the keys should be large to be visible and selectable **(Contradiction)**.

The invention predicts the next likely keys **(Principle-10: Prior Action)** and displays those keys with enlarged and bold font **(Principle-35: Parameter Change)**.

## 3. Summary and conclusion

A soft keyboard is a very special purpose keyboard. It has the advantage of entering data directly on the screen without having to struggle with a physical keyboard. On the other hand, the disadvantage of a soft keyboard is that it is quite slow to operate and not suitable for huge data entry purposes.

The software is popularly used in ATMs, Information Kiosks and other such places. As per the trend, the soft keyboard will be used more intensively in future days.

## 4. Reference

1. US Patent 5805157, Hexagonal shape soft keyboard for more accurate fingering., Bertram et al., assigned to IBM, Sep 98.

2. 5818451, Circular contact point for soft keyboard., Bertram et al., assigned to IBM, Oct 98

3. 5963671, "Enhancement of soft keyboard", invented by Comerford, et al., assigned to IBM in Oct 99

4. 5936614, "User defined keyboard entry system", Larry An et al., Assigned to IBM, Aug 99

5. 6573844, Predictive soft keyboard., Venolia et el., Assigned to Microsoft Corporation, June 2003

6. US Patent and Trademark Office (USPTO) site, http://www.uspto.gov/